\documentclass[preprint,aps, pra,showkeys,showpacs,preprintnumbers,amsmath,amssymb]{revtex4}
\usepackage{graphicx}
\usepackage{dcolumn}
\usepackage{bm}
\usepackage{color}

\begin{document}
\title{Revivals in the attractive BEC in a double-well potential \\and their decoherence}

\author{Krzysztof Paw\l owski}
\affiliation{%
Center for Theoretical Physics, Polish Academy of Sciences, \\
Al. Lotnik\'ow 32/46, PL-02-668 Warsaw, Poland
}
\affiliation{
Institute for Theoretical Physics, Warsaw University, \\
ul. Ho\.{z}a 69, PL-00-681, Warsaw, Poland
}
\author{Pawe\l~Zi\'n}%
\affiliation{Soltan Institute for Nuclear Studies, \\
ul. Ho\.{z}a 69, PL-00-681 Warsaw, Poland}%
\author{Kazimierz~Rza\.{z}ewski}%
\affiliation{%
Center for Theoretical Physics, Polish Academy of Sciences, \\
Al. Lotnik\'ow 32/46, PL-02-668 Warsaw, Poland
}
\affiliation{Faculty of Mathematics and Sciences, Cardinal Stefan Wyszy\'nski University, \\
ul. Dewajtis 5, PL-01-815, Warsaw, Poland}%
\author{Marek~Trippenbach}%
\affiliation{%
Institute for Theoretical Physics, Warsaw University, \\
ul. Ho\.{z}a 69, PL-00-681, Warsaw, Poland
}

\date{\today}

\begin{abstract}
We study the dynamics of ultracold attractive atoms in a weakly linked two potential wells. We consider an unbalanced initial state and monitor dynamics of the population difference between the two wells. The average imbalance between wells undergoes damped oscillations, like in a classical counterpart, but then it revives almost to the initial value. We explain in details the whole behavior using three different models of the system. Furthermore we investigate the sensitivity of the revivals on the decoherence caused by one- and three-body losses. We include the dissipative processes using appropriate master equations and solve them using the stochastic wave approximation method.
\end{abstract}

\pacs{03.75.Gg, 03.75.Kk, 67.85.De, 05.30.Jp}

\maketitle

\section{\label{sec:introduction}Introduction}

In 1962 Josephson predicted the current flow associated with coherent quantum tunneling of Cooper pairs through a barrier between two superconducting electrodes. This device is called a Josephson junction and the flow is proportional to the sine of the phase difference.
Jospehson also predicted that the time derivative of the phase difference is proportional to the voltage across the barrier. The corresponding relations can be found in \cite{josephson1962, jospehson1964, josephson1965}.
The first superfluid Josephson junctions were realized in superfluid $^3$He in 1997 \cite{pereverzev1997, backhaus1991, davis2002}.
The Josephson junction dynamics relies on the existence of two coupled macroscopic quantum states. 

With the advent of Bose Einstein condensates of weakly interacting gases a new experimental system has become available for the quantitative investigation of Josephson effects in a very well controlled environment \cite{javanainen1986}.
Note that the Josephson junction in this system consists of the two localized matter wave packets in the two wells coupled via tunneling of particles through the potential barrier. The authors of \cite{oberthaler2007} presented the experimental realization of the atomic Josephson Junction and compared the data obtained experimentally with predictions of a many-body two-mode model and a mean-field description.

In our paper we study two mode Bose-Hubbard model with the attractive interactions.
We focus on the dynamics of the state that is initially in a condensate state (all atoms occupying the same single particle state) with nonzero population imbalance.
During the time evolution there are Rabi oscillations in the mean value of the population imbalance.
These oscillations exhibit collapses and revivals, analogous to the well known phenomena in quantum optics (see Jaynes-Cummings model \cite{eberly}).
The revivals observed in quantum optics have been viewed as a proof of the existence of photons i.e. one could not explain this phenomenon within classical electromagnetic field interacting with atom.
In this study, we explain the mechanism of the revivals in the double-well system in the same manner as  revival predicted in James-Cummings model. However, we do not claim that the revivals are the evidence of particle existence. On the contrary, we reconstruct the revivals using only concepts from classical physics. Our classical model also predicts correctly collapses and Rabi oscillations.
Next we include in our consideration particle losses. Our numerical calculations prove that the revivals are extremely sensitive to decoherence -- loosing just a few particles destroys it almost completely.

The paper is organized as follows. In section \ref{sec:models} we introduce a two mode Bose Hubbard model. We focus on the attractive interaction case and shortly derive the continuous approximation.
This approximation maps the two mode Hamiltonian onto the Hamiltonian of a fictitious particle. We obtain a one-dimensional Schroedinger equation with an effective Planck constant $\hbar_{eff}$ inversely proportional to number of particles. The shape of the potential determines the dynamics of the system. This approach offers a better insight into the dynamics of the system. In section 
\ref{sec:times} we explain the origin of Rabi oscillations, their collapses and revivals. Furthermore we compute the timescales characteristic for these phenomena. The section \ref{sec:dec} is devoted to a  decoherence in the system and we discuss its influence on the dynamics. We include one- and three-body losses in the system and investigate their role in damping of revivals.
\section{\label{sec:models}Models}
We consider an evolution of the gaseous attractive Bose-Einstein condensate in a double-well potential in the two mode approximation.
In the limit of relatively weak interaction the system can be described by the Bose-Hubbard Hamiltonian
\begin{eqnarray}
\nonumber\hat {H}&=&\hbar\omega_0\hat{N} - \frac{J}{2}\left( \hat a_\mathbf{1}^{\dagger} \hat a_\mathbf{2} + \hat a_\mathbf{2}^{\dagger} \hat a_\mathbf{1}  \right)\\
& &-\frac{U}{2}\left( \left( \hat a_\mathbf{1}^{\dagger}\right)^2 \hat a_\mathbf{1}^2 + \left(\hat a_\mathbf{2}^{\dagger}\right)^{2} \hat a_\mathbf{2}^2 \right),
\label{eqn:ham}
\end{eqnarray}
where $J$ is the tunneling energy, $U$ is an average interaction energy per pair of atoms, $\hat{N}$ is a total number of particles and operators $\hat{a}_1 ^{\dagger}$ and $\hat{a}_2 ^{\dagger}$ create one atom in left and right well respectively.
In all cases except these with particle losses, it is useful to omit constant terms in \eqref{eqn:ham} and rewrite it in the dimensionless form
\begin{equation}
\hat {H}_2 = - \left( \hat a_\mathbf{1}^{\dagger} \hat a_\mathbf{2} + \hat a_\mathbf{2}^{\dagger} \hat a_\mathbf{1}  \right)
- \frac{\gamma}{2} \left( \hat a_\mathbf{1}^{\dagger} \hat a_\mathbf{1} - \hat a_\mathbf{2}^{\dagger} \hat a_\mathbf{2} \right)^2,
\label{eqn:ham2}
\end{equation}
where $\gamma=\frac{UN}{J}$. This form of Bose-Hubbard Hamiltonian determines our dimensionless time unit $\tau :=\frac{J t}{2\hbar}$.
To get a better insight in the structure of the Hamiltonian (more 'intuitive one')
one can use a continuous approximation, valid for large $N$.
Consider the expansion of the state of system in the Fock basis: $|\psi\rangle=\sum_{n=0}^N \psi_n |n, N-n\rangle$, where the ket $|n, N-n\rangle$ means a state with $n$ atoms in the 'right' well and rest of them in the 'left' well. The Schr\"odinger equations for the coefficients $\psi_n$ can be written as
\begin{eqnarray}
\nonumber \dot{\imath}  \frac{d \psi_n}{d\tau} &=- \frac{\gamma}{2N} \left( 2n- N \right) ^2 \psi_{n} -\sqrt{n\left( N-n+1 \right)} \psi_{n-1} \\
 &- \sqrt{\left( n+1 \right) \left( N-n \right) }  \psi_{n+1}.
\label{eqn:coeff}
\end{eqnarray}
Assuming that $N \gg 1$ and  $\psi_n$ are negligible near the 'boundary' (for $n$ close to zero and $N$) we can approximate equation \eqref{eqn:coeff} getting
\begin{equation}
 \dot{\imath}\frac{1}{N}\frac{d \psi_n}{d\tau} = -\frac{\sqrt{1-z_n ^{2}}}{2}\left( \psi_{n-1} + \psi_{n+1}\right) - \frac{\gamma}{2}z_n ^2 \psi_{n},
\label{eqn:help1}
\end{equation}
where $z_n =\frac{(N-n)-n}{N}$.
In the same limit a discrete set of the coefficients $\psi_n$ may be replaced by a 'wave function' $\psi(z)$. Upon replacing $ (\psi_{n-1} + \psi_{n+1} - 2 \psi_n) \simeq \frac{4}{N^2}\frac{\partial^2}{\partial z^2} \psi(z)$ the equation (\ref{eqn:help1}) takes the form 
\begin{eqnarray}
   \dot{\imath} \frac{1}{N} \frac{\partial \psi}{\partial \tau}=-\frac{2\sqrt{1-z^2}}{N^2} \frac{\partial ^2 \psi}{\partial z^2}  + V\left(z\right) \psi ,
 \label{ciaglyham0}
\end{eqnarray}
where
\begin{equation}
V(z)=-\sqrt{1-z^2}-\frac{\gamma}{2}z^2.
\label{eqn:potential}
\end{equation}
We further simplify the equation \eqref{ciaglyham0} through omitting the term $\sqrt{1-z^2}$ in front of the second derivative:
\begin{eqnarray}
   \dot{\imath} \frac{1}{N} \frac{\partial \psi}{\partial \tau}=-\frac{2}{N^2} \frac{\partial ^2 \psi}{\partial z^2}  + V\left(z\right) \psi .
 \label{ciaglyham}
\end{eqnarray}
The latter equation describes a fictitious particle of mass $m_{eff}=\frac{1}{4}$ with the effective Planck constant  $\hbar_{eff} = \frac{1}{N}$.
We will consider only evolution limited to small mean value of position $\langle z\rangle$. In these cases replacing the effective mass $\frac{1}{4\sqrt{1-z^2}}$ from Eqn. \ref{ciaglyham0} with only $m_{eff}=\frac{1}{4}$ will not change qualitatively evolution. 
\begin{figure}
\centering
\includegraphics[width=8.6cm,clip]{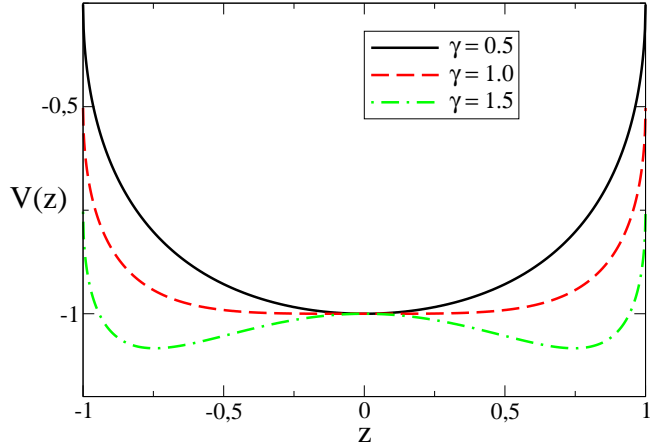}
\caption{(Color online) Shapes of effective potential for various values of $\gamma$. }
 \label{fig:potential}
\end{figure}
The shape of the effective potential is shown in Fig. \ref{fig:potential}. With increasing $\gamma$  the shape of the effective potential \eqref{eqn:potential} turns from a single minimum located at $z=0$, to a double-well structure for $\gamma>1$.

\begin{figure}
\centering
\includegraphics[width=8.6cm, clip]{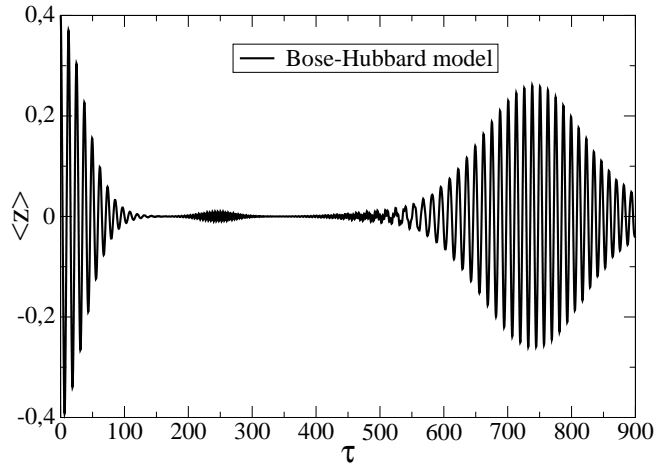}
\caption{ Numerical solution of Bose-Hubbard model. Average relative atoms' imbalance between wells $\langle z\rangle$ for $1000$ particles, the initial average imbalance $z_0=0.4$, the initial phase $\phi_0=0$ and $\gamma=1.0$.}
 \label{fig:exact}
\end{figure}
Let us define the family of initial states, which we will consider in the rest of the paper. As an initial state we take a coherent state with $N$-atoms in the same single-atom state
\begin{equation}
\vert \psi(0) \rangle =\frac{1}{\sqrt{N!}} \left( \sqrt{\frac{1+z_0} {2} } a_1^{\dagger} + e^{\dot{\imath} \phi_0} \sqrt{\frac{1-z_0}{2} } a_2^{\dagger} \right) ^N \vert 0\rangle .
\label{eqn:stpoczatkowy}
\end{equation}
The $z_0$ is the initial imbalance 
\begin{equation}
 z_0 = \langle \psi(0)| \hat z |\psi(0) \rangle,
\end{equation}
where
\begin{equation}
\hat{z}:=(a_1^{\dagger}a_1-a_2^{\dagger}a_2)/N
\end{equation}
is a population imbalance operator. The parameter $\phi_0$ can be interpreted as a relative phase between two wells. 

In the continuous approximation the initial state corresponding to the state \eqref{eqn:stpoczatkowy} can be constructed upon it's expansion in  the Fock basis
\begin{equation}
|\psi\rangle = \sum_{n=0}^N \sqrt{\binom{N}{n}} p^{n/2} q^{(N-n)/2} e^{\dot{\imath} (N-n)\phi_0}|n, N-n\rangle ,
\label{eqn:statetemp}
\end{equation}
where $p=\left(\frac{1+z_0}{2}\right)$ and $q=1-p$. The square of modulus of the expansion coefficients in \eqref{eqn:statetemp} is just a probability function for the binomial distribution. In the continuous limit ( $N\rightarrow \infty$ and $z_n\rightarrow z$) they converge to a Gaussian function and we get
\begin{eqnarray}
 \psi \left( z\right) &=& \sqrt[4]{\frac{N}{2 \pi \left(1 - z_0 ^{2} \right)}}\;e^{-\frac{N \left( z - z_0 \right) ^{2}}{4 \left(1 - z_0 ^{2}\right)}} e^{\dot{\imath} N\frac{1-z}{2}\phi_0}. \label{eqn:initial}
\end{eqnarray}
The above discussion is presented in more details in \cite{schchesnovich2007, zin04_2008}.

The continuous model together with the initial state proposed above give us a clear picture of what will happen in the system. We have a Gaussian wave packet moving in the potential $V(z)$. The shape of the potential will be crucial for the dynamics. 
In all further considerations we focus on the case $\gamma=1$, where in the effective potential the harmonic terms $z^2$ are canceling and we approximate it by the first two leading terms:
\begin{equation}
V_{eff}^{\gamma=1} \approx -1+\frac{z^4}{8}.
\label{eqn:eff_pot_g1}
\end{equation}
Having chosen $\gamma$ and the initial state $|\psi(0) \rangle$ the only free parameter is the number of particles $N$.

At last, let us focus on the dynamics. 
We evolve the state $| \psi(t) \rangle$ using the Schroedinger equation (\ref{eqn:coeff}). Fig. \ref{fig:exact} shows the $ \langle \psi(t)| \hat z |\psi(t) \rangle $ in the case $N=500$. Here we observe Rabi oscillations, with distinct collapse at time $\tau\approx 150$ and the revival at $\tau\approx 750$. 

In the next sections we focus our attention on the variable $\langle \psi(t)| \hat z |\psi(t) \rangle$.

\section{\label{sec:times} The origin of Rabi oscillations, collapses and revivals.}
\subsection{\label{subsec:rabi}Rabi oscillations}
\begin{figure}
 \centering
 \includegraphics[width=8.6cm]{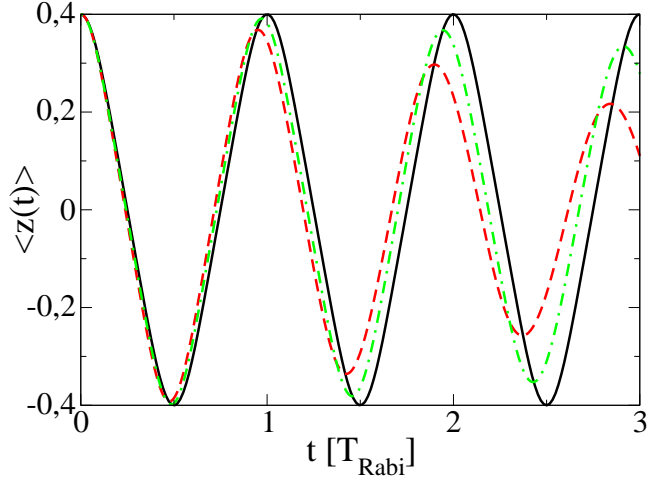}
 \caption{(Color online) Comparison between evolution of $\langle z\rangle$ given by Bose-Hubbard model for 2000 particles (red - dashed line), for $8000$ particles (green, dot-dashed line) and classical model (black, solid line) for $z_0=0.4$, $\phi_0=0$ and $\gamma=1$. The scale on x-axis corresponds to analytical calculated time of the Rabi oscillation.}
 \label{fig:classical}
\end{figure}
Due to the tunneling between the wells for most initial conditions $\langle \hat{z}\rangle $ will oscillate periodically in time. This phenomena can be easily illustrated within the
classical field theory (substituting the annihilation and creation operators by  complex numbers).
This approximation may be understood as a transition from the Schr\"odinger equation (\ref{ciaglyham}) to the Newton equation
 \begin{equation}
m_{eff} \frac{d^2 z}{d\tau ^2} = - \frac{\partial V}{\partial z}
\label{eqn:newton}
\end{equation}
The motion of the fictitious particle described by the above equation is periodic. In the spirit of quantum optics we call these the Rabi oscillations.
Using the Newton equation we calculate their period $T_{Rabi}$. The formula following from Eq. \eqref{eqn:newton} reads
\begin{equation}
T_{Rabi} = 2\int_{z_{min}}^{z_{max}}\frac{dz}{\sqrt{2\left(E-V(z)\right)/m_{eff}}}
\label{eqn:rab}
\end{equation}
The energy $E$ is just a sum of a potential and a kinetic energy $\frac{m_{eff}}{2}\frac{d^2z}{dt^2}$. From the initial state it's clear that $E = V(z_0) \approx -1+\frac{z_0^4}{8}$. Substituting this value into \eqref{eqn:rab} we can estimate
\begin{equation}
T_{Rabi} = \frac{ 4\sqrt{\pi}\Gamma \left( \frac{5}{4}\right)}{z_0 \Gamma \left( \frac{3}{4}\right) } .
\end{equation}
In Fig. \ref{fig:classical} we compare this result with numerical simulation of Bose-Hubbard model for a short time evolution getting a good qualitative agreement. From this figure we can also conclude,  that the approximation become accurate in the limit $N\rightarrow\infty$.

\subsection{\label{subsec:wigner}Collapse}

The single trajectory of Newton equation contains oscillations which continue forever, so 
it does not describe the phenomenon of a collapse. 
We argue that one can obtain the collapse using Newton equation with a set of random initial conditions, with appropriate probability distribution.Our method will be similar to a semiclassical model exploited in \cite{micheli2003}.
This approach is also equivalent to a stochastic classical field theory

\begin{figure}
\centering
\includegraphics[width=8.6cm]{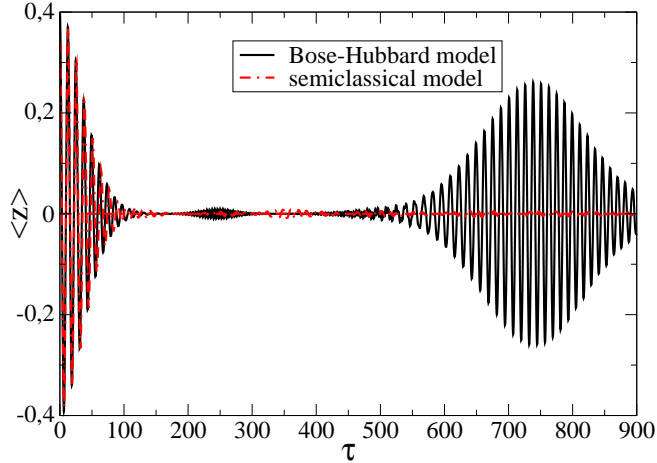}
\caption{(Color online) Comparison between evolution of $\langle z\rangle$ given by Bose-Hubbard model (black - solid line) and semiclassical model (red - dashed line).}
\label{fig:wigner}
\end{figure}

Central issue is: is there an appropriate probability distribution corresponding to the initial state \eqref{eqn:initial}. One of the candidates is the Wigner distribution if it is positive.
Since the classical equation we use is the Newton equation corresponding to the Schr\"odinger equation \eqref{ciaglyham} the Wigner function can be calculated as
\begin{eqnarray}
\nonumber\mathcal{W}\left(z,\phi\right) &=& \frac{N}{4\pi}\int \mbox{d}\lambda \psi^{\ast}( z-\frac{\lambda}{2})\psi( z+\frac{\lambda}{2}) e^{-\dot{\imath}\lambda N \phi/2}=\\
& &=\frac{N}{2\pi} e^{-\frac{N\left(z-z_0\right)^2}{2\left( 1-z_0^2\right)}} e^{-N\frac{\left( 1-z_0^2\right)(\phi + \phi_0)^2}{2}}.
\label{eqn:wignerdist}
\end{eqnarray}
The procedure used to describe the dynamics will consist of the following steps. 
We draw a set of (in our case $50 000$) initial conditions, propagate them for some interval of time $t$ and form the distribution again. Using this new distribution we can calculate any expectation value. In particular in Fig. \ref{fig:wigner} we present the evolution of $\langle z \rangle$ obtained this way and compare it with the numerical solution of the quantum Bose Hubbard model. The agreement is satisfactory and gives a correct prediction of the collapse. It indicates that the collapse is a result of initial conditions uncertainty (spread). 
We clearly see that the revivals are not present in this approach.
This is consistent with the understanding of this phenomena in quantum optics. 

\begin{figure}
\begin{minipage}[t]{0.21\textwidth}
\centering
(a) $\tau=0$

\includegraphics[width= \textwidth, clip]{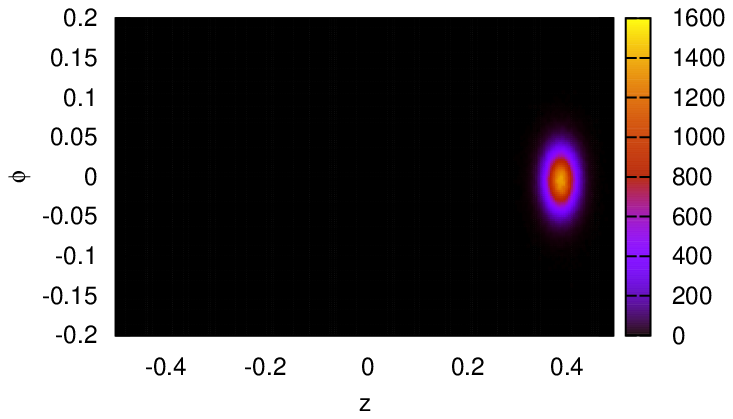}
\end{minipage}
\hspace{0.05\textwidth}
\begin{minipage}[t]{0.21\textwidth}
\centering
(b) $\tau=40$

\includegraphics[width= \textwidth, clip]{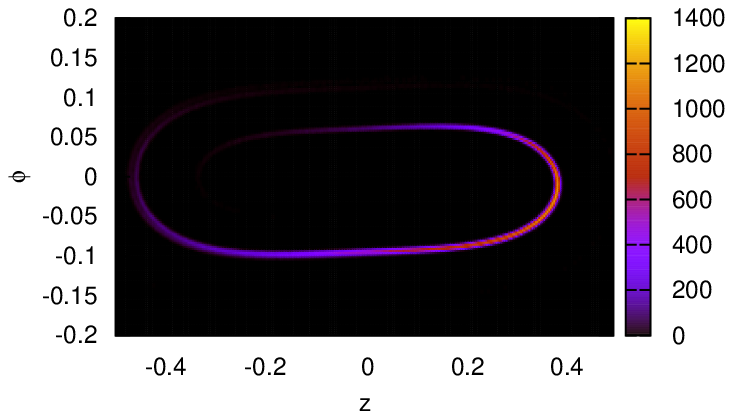}
\end{minipage}

\begin{minipage}[b]{0.21\textwidth}
\centering
(c) $\tau=80$

\includegraphics[width= \textwidth, clip]{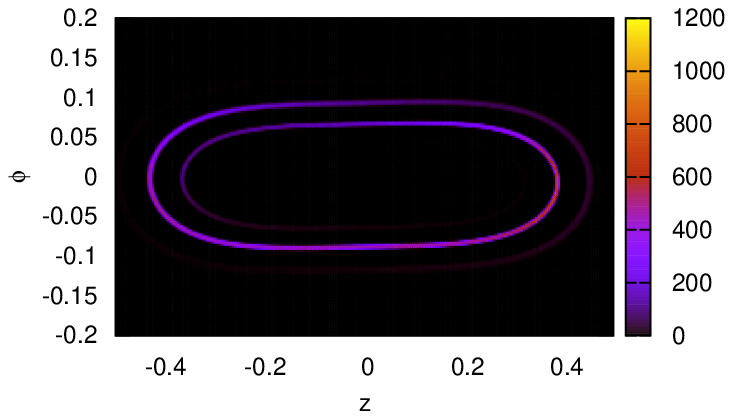}
\end{minipage}
\hspace{0.05\textwidth}
\begin{minipage}[b]{0.21\textwidth}
\centering
(d) $\tau=160$

\includegraphics[width= \textwidth, clip]{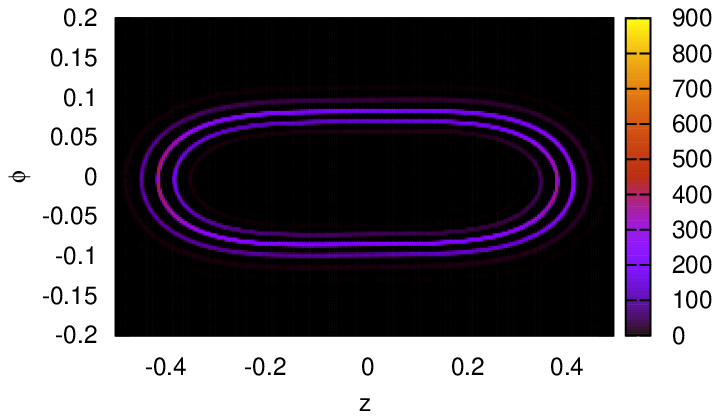}
\end{minipage}
\caption{(Color online) The evolution of classical distribution in $z-\phi$ space, initially equal to the Wigner function. The bright regions corresponds to high probability of finding an atom in this region. Parameters in simulations $N=2000$, $z0=0.4$, $\phi_0=0.0$ and $\gamma=1$. }
\label{fig:wignerzp}
\end{figure}
In Fig. \ref{fig:wignerzp} we present the reconstruction of the classical distribution in the phase space $z-\phi$ at different stages of the evolution. This picture illustrates generation of the damping of Rabi oscillations.
When the distribution reaches a shape of ring, the average $\langle z\rangle$ becomes $0$ and the collapse occurs. The merging of the distribution's tails is due to the dispersion of energy in the initial conditions
\begin{equation}\label{eqn:disp}
\Delta E = \sqrt{ \langle  \hat H^2 \rangle - \langle  \hat H  \rangle^2} = 1/\sqrt{N} +O(\frac{1}{N}),
\end{equation}
and the dependence between the time of one period and the energy $T_{Rabi}\propto E^{-1/4}$ (see Eq. \eqref{eqn:rab}).
Using \eqref{eqn:disp} and \eqref{eqn:rab} one can show that 
\[
T_{col}\propto \sqrt{N}.
\]
\begin{figure}
 \includegraphics[width=8.6cm, clip]{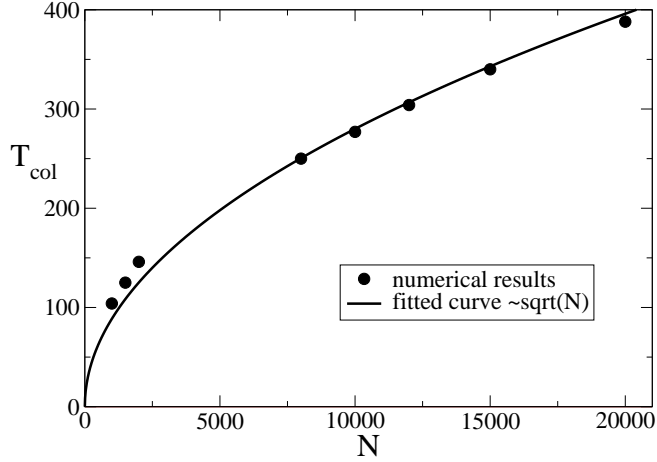}
\caption{Timescales of collapses for different number of atoms $N$ -- results of numerical simulation of Bose-Hubbard model (black circles). As predicted in the text the dependence is like $\sqrt{N}$ (solid line -- fitted curve).}
\label{czaskolapsow}
\end{figure}
In Fig. \ref{czaskolapsow} we compare our prediction with the numerical results getting a good agreement.
The derivation presented above is only a very rough estimation of the real process. In fact the collapse occurs when the distribution 'winds up' three or four times so when the distribution  smears uniformly over ellipses in the phase space (Fig. \ref{fig:wignerzp}). However we think that the presented mechanism of collapse is intuitive and useful for the qualitative predictions.
\subsection{\label{subsec:revival}Time of revival}
\begin{figure}
\centering
\includegraphics[width=8.6cm, clip]{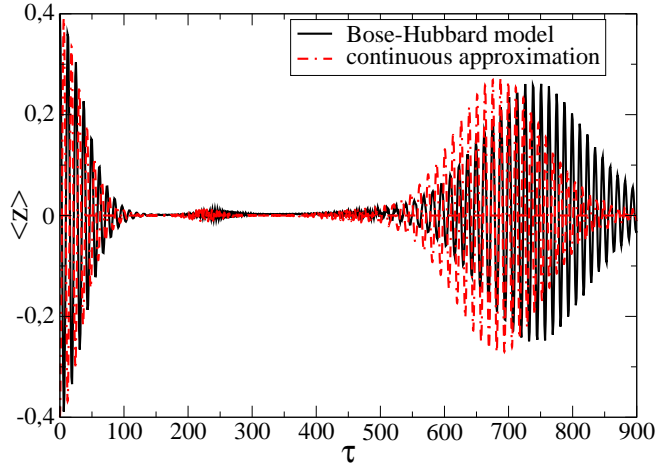}
\caption{(Color online) Comparison between the numerical solution of Bose-Hubbard model (solid, black line) and Schr\"odinger like equation (dashed, red line) for $N=2000$ particles and initial imbalance $z_0=0.4$.}
 \label{fig:contin}
\end{figure}
The most famous revival is the one in the evolution of a two level atom interacting with state of a single electromagnetic mode. This was first predicted in \cite{eberly}, and finally observed \cite{haroche1996}. The theoretical considerations used to calculate the revival time in \cite{eberly} may be also applied to our system. 
However, to follow the derivation we need the structure of energy levels. 

On the other hand we see in Fig. \ref{fig:contin} the Schr\"odinger equation reveals the revivals. The time of revival in the frame of Schr\"odinger equation agree qualitatively with $T_{rev}$ from Bose-Hubbard model. Thus we compute the time approximating the system with the Schr\"odinger equation. In this way we gain the knowledge of energy levels, which for particle moving in a smooth potential may be easily estimated from WKB theory

We decompose the state of the system $|\psi(\tau)\rangle$ in the basis of the eigenstates
\begin{equation}
 \mid\psi(\tau)\rangle = \Sigma _{n} A_n (0 )e^{\dot{\imath} E_n \tau /h_{eff} }\mid E_n \rangle\;\mathrm{,}
\end{equation}
where $| E_n \rangle$ denotes the $n^{th}$ eigenstate of the Hamiltonian \eqref{ciaglyham}. 
Thus, the evolution of $\langle \hat{z}\rangle$ is given by the equation:
\begin{equation}
 \langle \hat{z}\rangle (\tau )=\Sigma _{n, m} A_n(0) A_m ^{\ast}(0) e^{-\dot{\imath}\left( E_{n} -E_{m}\right)\tau /h_{eff}} \langle E_m \mid \hat z \mid E_n \rangle .
\label{eqsum}
\end{equation}

Here, we estimate the energy spectrum using the Wentzl-Kramers-Brillouin approximation (WKB, see for instance \cite{griffiths2004}). 
The stationary version of the Schr\"odinger equation \eqref{ciaglyham} has a form:
\begin{eqnarray}
 \frac{\partial ^2 \psi_n}{\partial z^2}  + \kappa^{2}(z)\psi_n =0 ,
\end{eqnarray}
where $ \kappa(z) =  \sqrt{ 2 m_{eff}\frac{E_n - V(z)}{\hbar_{eff}^2}}$.
One of conclusions of the WKB approximation is the following relation
\begin{equation}
\int_{z_{max}} ^{z_{min}} \kappa (z)\,\mbox{d}z = \left( n+\frac{1}{2}\right)\pi ,
\label{eqn:wkb}
\end{equation}
which is widely used to estimate the eigenenergies $E_n$. The points $z_min$ and $a_max$ are the turning points for a particle with energy $E_n$ moving in the potential $V(z)$.
The integral in \eqref{eqn:wkb} has an analytical solution, 
\begin{displaymath}
\int_{z_{max}} ^{z_{min}} \kappa (z)\,\mbox{d}z = \frac{N}{16} \frac{\left(\tilde{E}_n\right)^{3/4}\sqrt{\pi}\Gamma \left(1/4\right)}{\Gamma \left(7/4\right)},
\end{displaymath}
where $\tilde{E}_n=8 (E_n+1)$. Hence we approximate the eigenenergies using the following formula:
\begin{equation}
 \tilde{E}_n =  \left[\frac{D}{N}\left( 2n+1\right)\right]^{\frac{4}{3}}.
\label{eqnpoziom}
\end{equation}
 Here $D=\frac{8 \sqrt{\pi} \Gamma \left(7/4\right)}{\Gamma \left(1/4\right)}$. We define the number $\bar{n}$, which corresponds to the index of state with the energy equal to the average energy in the system. It is given by
\begin{equation}
\bar{n} =\frac{N}{2 D}z_0^3-\frac{1}{2}
\label{eqn:avern}
\end{equation}
The investigated sum \eqref{eqsum} will be maximal if all the components have the same phase. Obviously this condition couldn't be fulfilled for all terms in \eqref{eqsum}, but it is sufficient to ensure it for major terms. So let us investigate the differences between phases for terms with indices close to $\bar{n}$. Thus, we are looking for a time $T_{rev}$ that satisfies the relation:
\begin{equation}
T_{rev} \left(E_{\bar{n}+2}-2E_{\bar{n}+1}+E_{\bar{n}}\right) = 2\pi \hbar_{eff}.
\label{eqn:reviv}
\end{equation}
After substituting equations \eqref{eqn:avern} and \eqref{eqnpoziom} to the condition \eqref{eqn:reviv} we get the following estimation for the revival  time
\begin{equation}
 T_{rev} = \left(\frac{9z_0^2\pi}{D^2}\right)^2N.
\end{equation}

\begin{figure}
 \includegraphics[width=8.6cm, clip]{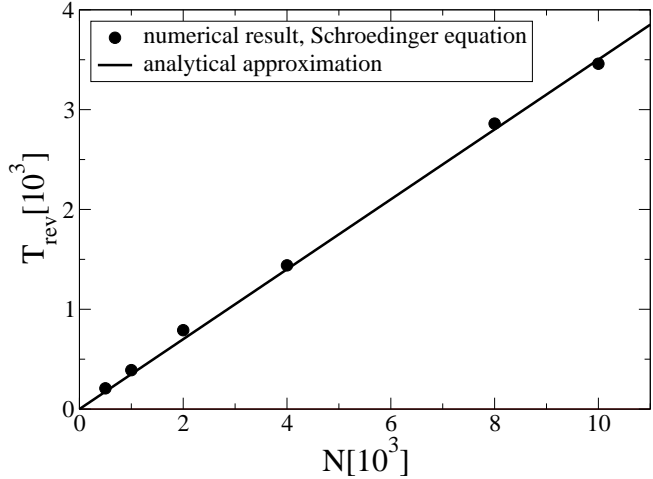}
\caption{Times of revival for different number of atoms $N$. Dots -- the numerical simulation of the Schr\"odinger equation ($z_0=0.4, \gamma=1$), solid line -- the analytical estimation.}
\label{fig:revivals}
\end{figure}
The test of the linear dependence $T_{rev}$ with respect to the number of particles $N$ is presented in Fig. \ref{fig:revivals}.

\subsection{\label{subsec:revclas}Classical revival}
Here we show, that though we start from the state that entangles atoms from both potential wells \eqref{eqn:stpoczatkowy} 
the main features (including revivals) of the quantity $\langle z\rangle (t)$ can be reconstructed classically. More specifically we find some classical distribution in the $z-\phi$ space and we use it to distribute initial conditions fot the Newton equation \eqref{eqn:newton}. After averaging these classical evolutions we get some average trajectory  $\langle z\rangle (t)$. Here, we present our method in the system of $N=2000$ atoms and initial imbalance $z_0=0.4$.

From Fig. \ref{fig:exact} we see that the revival occurs about $\tau=740$. We will find such initial conditions for Newton equation, that the revival will be enforced. One way  is to choose such discrete set $z_0^{(n)}$ that $T_{rev}$ will be a multiple of the Rabi periods $T_{Rabi}(z_0^{(n)})$:
\begin{equation}
\frac{T_{rev}}{T_{Rabi}(z_0^{(n)})}=n\, \in\, N\quad \Rightarrow\quad z_0^{(n)} = \frac{4\sqrt{\pi } n \Gamma \left(\frac{5}{4}\right)}{T_{rev} \Gamma \left(\frac{3}{4}\right)} .
\label{eqn:z0n}
\end{equation}
Thus when we evolve the point $ z_0^{(n)}$ using the Netwon equation, at time $T_{rev}$ it has to return to its initial position. However, to satisfactory reconstruct the mean value $\langle z\rangle$ we have to compute average of classical trajectories using some weights. We propose weights based on the Wigner distribution:
\begin{displaymath}
w(n) = \int_{\frac{z_0^{(n)}+z_0^{(n-1)}}{2}} ^{\frac{z_0^{(n)}+z_0^{(n+1)}}{2}} \int_{-\infty}^{\infty} \mathcal{W}\left(z,\phi\right) \mbox{d}\phi\,\mbox{d}z
\end{displaymath}
Thus our final approximation is 
\begin{equation}
z(t) = \sum_{n=0}^{\infty}w(n)\,z^{(n)}(t),
\label{eqn:zt}
\end{equation}
where $z^{(n)}(t)$ is the solution of Newton equation with initial conditions $z_0=z_0^{(n)}$ defined in \eqref{eqn:z0n} and $\frac{dz}{dt}\mid_{t=0}=0$.
\begin{figure}
\includegraphics[width=8.6cm, clip]{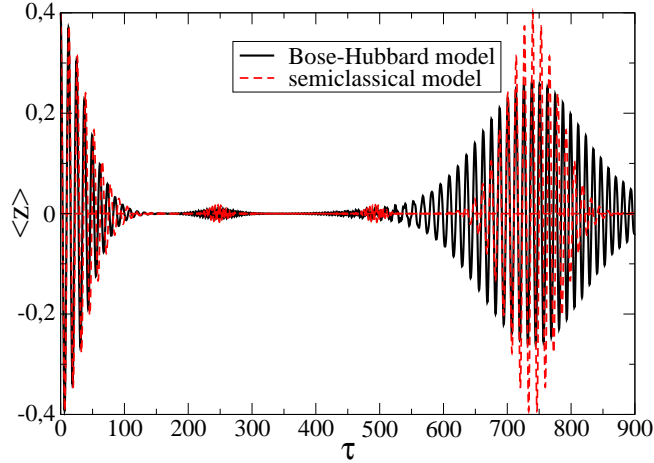}
\caption{(Color online) Comparison between the numerical solution of Bose-Hubbard model (solid, black line) and average over classical trajectories(dashed, red line) for $2000$ particles.}
 \label{fig:revclas}
\end{figure}
The comparison between our approximation \eqref{eqn:zt} and the numerical solution of the Bose-Hubbard model is presented in Fig. \ref{fig:revclas}. Such constructed solution \eqref{eqn:zt} reveals not only the times of revivals, but also the time of collapse and the period of Rabi oscillation.

\section{\label{sec:dec}Decoherence}
So far most experiments with BEC in a double-well potential were performed with atoms with repulsive interactions \cite{oberthaler2007, albiez2004}. On the other hand experimental data can be collected over very long time, up to a few seconds. Assuming that the similar experiments will be possible in the near future, what will be the biggest challenge on the way to observe revivals? We found that one of the crucial parameters is the number of atoms. Hence we think that it is very instructive and important to study our system in the context of the particle loss.

In this paper we demonstrate two mechanisms of the decoherence. The first of them is due to collisions with a hot background gas. A typical effect of such a collision is a loss of a cold atom from the trap. The second process under consideration is a three-body loss due to a recombination. If three condensed atoms meet, two of them may create a bound state and push the third to fly away carrying the energy gained from bound atoms.
In the next subsections we review some of the main features of these decoherence sources and present master equations which cover both the free evolution of BEC and atoms' losses. These master equations are of the type
\begin{equation}
\frac{d\hat{\rho}}{dt}=\frac{\dot{\imath}}{\hbar} \left[ \hat{\rho}, \hat{H}_0\right] + L\hat{\rho},
\end{equation}
where $\hat{H}_0$ is the Hamiltonian of a quantum subsystem and the superoperator $L$ has the form:
\begin{equation}
L\hat{\rho} = \sum_k \left(2\hat{C}_k^\dagger \hat{\rho} \hat{C}_k -\hat{C}_k^\dagger \hat{C}_k\hat{\rho}-\hat{\rho}\hat{C}_k^\dagger \hat{C}_k \right).
\end{equation}
We solve the equations numerically in the frame of the quantum trajectory approximation (also called the stochastic wave approximation) described in details in \cite{moelmer1993, carmichael1991}. In this approximation the density operator of the system is written as the average of the stochastic wave functions
\begin{equation}
\hat{\rho}(t)=\lim_{M\rightarrow\infty}\frac{1}{M}\sum_{m=1}^{M} |\psi^m_{stoch}(t)\rangle\langle\psi^m_{stoch}(t)| ,
\end{equation}
where $|\psi^m_{stoch}(t)\rangle$ is a $m$-th stochastic wave function at the time $t$. The function $|\psi^m_{stoch}(t)\rangle$ is also called a single quantum trajectory. It is obtained in the stochastic evolution described in the following.

In the evolution of the system during some small interval $\Delta t$ one of the elementary processes occurs: either atoms undergo the counterpart of free evolution or one of the quantum jumps.

The quantum jumps are just actions of the operators $\hat{C}_k$ on the state of the system
\begin{equation}
|\psi_{stoch}\rangle\mapsto \hat{C}_k |\psi_{stoch}\rangle .
\end{equation}
The $k$-th quantum jumps occurs with  the probability $p_k=\langle \hat{C}_k^{\dagger} \hat{C}_k \rangle \Delta t$, where the average $\langle \hat{C}_k^{\dagger} \hat{C}_k \rangle$ is computed for the stochastic wave and generally it differs between trajectories.

The 'free evolution' is represented by the effective operator
\begin{equation}
\tilde{H} = \hat{H} - \frac{\dot{\imath} \hbar}{2}\hat{C}_k^{\dagger} \hat{C}_k.
\label{eqn:heffree}
\end{equation}
If none of the quantum jumps occur (what happens with the probability $1-\sum_k p_k$) the system undergoes the transformation
\begin{equation}
|\psi_{stoch}\rangle\mapsto e^{-\frac{\dot{\imath} \tilde{H}}{\hbar}\Delta t}|\psi_{stoch}\rangle .
\end{equation}

The operators $\tilde{H}$ and $\hat{C}_k$ are non-hermitian, so the stochastic wave function must be normalized after every time step $\Delta t$.

A single stochastic trajectory is sometimes interpreted as a single experimental realization of the quantum evolution. This was the case of the famous experiments showing the quantum jumps \cite{Dehmelt1986}. The stochastic evolution was successfully applied in the study of the decoherence in BEC \cite{sinatra1998, yun2008} and the interference of two Fock states \cite{chough1997}. Recently, the quantum jumps trajectories (generated in the frame of the large deviation method) turned out to be very useful for studying the phase transitions \cite{andrieux2010}.

\subsection{\label{subsec:1body}One-body losses}

Due to non-perfect vacuum all experiments with ultracold atoms are performed in the presence of a background residual gas. Atoms from the residual gas affect the condensate via collisions with ultracold atoms. We assume that the typical energy of an incoming atom from the background is so high that during the collision the condensed atom is kicked out from the double-well trap. This kind of losses plays the major role for long lasting experiments, which may be the case of presented revivals. We propose the following phenomenological master equation which includes the one-body losses
\begin{equation}
\partial _t\hat{\rho }=-\frac{i}{\hbar }\left[\hat{H}, \hat{\rho }\right]+\kappa_1 \sum_{k=1}^2 \left( \left[ \hat{a}_k, \hat{\rho } \hat{a}_ {k}^{\dagger }\right]+\left[ \hat{a}_k \hat{\rho}, \hat{a}_ {k}^{\dagger }\right] \right),
\label{eqn:master1}
\end{equation}
where $\hat{H}$ is given by equation \eqref{eqn:ham} and $\kappa_1$ is a rate of one-body losses.
In a case of losses caused by the residual gas, the parameter $\kappa_1$ is just an inverse of the lifetime in a trap \cite{pawlowski2010}. The form of equation \eqref{eqn:master1} has been proven in the case of suppressed tunneling ($J=0$) in Ref. \cite{sinatra1998}.

In the case of the master equation \eqref{eqn:master1} the evolution operator and quantum jump operators have the form
\begin{eqnarray}
\nonumber\tilde{H}&=& \hat{H}\\
\nonumber\hat{C}_1&=& \sqrt{2\kappa_1} \hat{a}_1\\
\hat{C}_2&=& \sqrt{2\kappa_1} \hat{a}_2.
\label{eqn:jumps1}
\end{eqnarray}
\begin{figure}
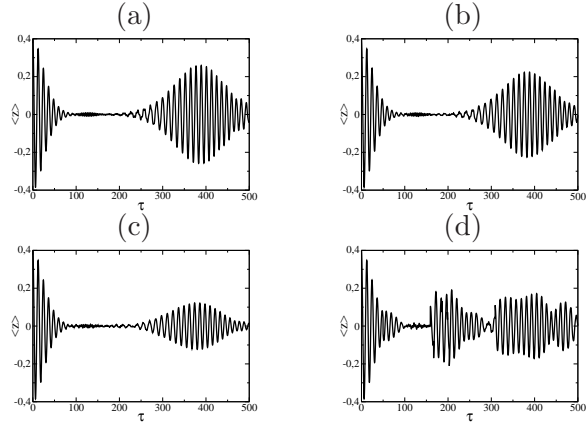

\begin{minipage}[t]{0.20\textwidth}
\centering
(a)

\includegraphics[width= \textwidth, clip]{exact3}
\end{minipage}
\hspace{0.05\textwidth}
\begin{minipage}[t]{0.20\textwidth}
\centering
(b)

\includegraphics[width= \textwidth, clip]{z1_1body}
\end{minipage}

\begin{minipage}[b]{0.20\textwidth}
\centering
(c)

\includegraphics[width= \textwidth, clip]{z4_1body}
\end{minipage}
\hspace{0.05\textwidth}
\begin{minipage}[b]{0.20\textwidth}
\centering
(d)

\includegraphics[width= \textwidth, clip]{z1_1body_single}
\end{minipage}
\caption{ Evolution of $\langle z \rangle (\tau)$ (a) without losses, (b) with in average one quantum jump before revival, (c) with in average four quantum jumps before revival. (d) Single realization of stochastic wave approximation. The phase of Fock components of stochastic wave change randomly. This process leads to jumps of average $\langle z\rangle$, which significantly differs from its average evolution. In the all figures $z_0=0.4$, the initial total number of atoms $N=1000$ and $\gamma=1$.}
\label{fig:1body}
\end{figure}
Note that the free evolution operator $\tilde{H}$ is given just by the hermitian operator $\hat{H}$. According to the equation \eqref{eqn:heffree} this operator should have another form, namely
\begin{displaymath}
\tilde{H}_{1}=\hat{H}-\dot{\imath} \hbar \kappa_1 (\hat{a}_1^\dagger \hat{a}_1+\hat{a}_2^\dagger\hat{a}_2)=\hat{H}-\dot{\imath} \hbar \kappa_1\hat{N},
\end{displaymath}
where $\hat{N}$ is the operator of the total number of atoms in both wells. Then, after the nonhermitian evolution
\begin{displaymath}
 |\psi_{stoch}\rangle (\tau+\Delta \tau) = e^{-\frac{\dot{\imath} \tilde{H}_1}{\hbar}\Delta \tau} |\psi_{stoch}\rangle (\tau) ,
\end{displaymath}
the stochastic wave function $|\psi_{stoch}\rangle (\tau+\Delta \tau )$ ought to be normalized. However, in this case the normalization would just mean division the state by $\exp(-N \kappa_1 \Delta \tau)$, where $N$ would be the total temporary number of atoms. Thus the procedure of evolution with the non-hermitian operator and then the state's normalization is equivalent to the free hermitian evolution, resulting from operator $\hat{H}$ defined in Eqn. \eqref{eqn:ham}.

We carried out the series of stochastic evolutions for different damping parameters $\kappa_1$. In Fig. \ref{fig:1body} we show the evolutions with $\kappa_1=0$ (Fig. \ref{fig:1body}(a), with such  damping that in the system up to the first revival one particle is lost (Fig. \ref{fig:1body}(b) and  four particles are lost (Fig. \ref{fig:1body}c). We obtained these results averaging over only $100$ quantum trajectories.
Figures \ref{fig:1body}(d) displays just a single quantum trajectory. We see a fast damping of revivals, even if the loss of the total number of atoms is relatively small. Even for just four lost atoms, the amplitude of revival decreases more than twice. It gives us the following qualitative criterion for the upper value of parameter $\kappa_1$,
\begin{displaymath}
\kappa_1 N t_{rev}<1 .
\end{displaymath}
Assuming number of atoms $N\sim 1000$ and $t_{rev}\sim 50\mu s$ (estimated from typical parameters of double-well potentials) we get $\kappa_1<20/$s. In typical experiments with BEC this parameter is two orders of magnitude smaller \cite{pawlowski2010}. Thus, the mechanism of one-body losses shouldn't prevent the revivals. On the other hand we haven't included other sources of one-body losses, like for instance due to heating.

One may observe, that the quantum trajectories (Fig. \ref{fig:1body}(d) differ significantly from the result of averaging. Even though in a single stochastic realization the amplitudes of revival are typically as large as in the case without losses. However, the decrease of revivals' amplitude in average is fairly intuitive. Let us remind, that the time of revival is proportional to the number of atoms. If we permit atomic losses the number of particles may change so the time of revival occurrence become shorter. The average $\langle z\rangle$ has solution oscillating in time with period $T_{Rabi}$. Thus the averaging over many stochastic trajectories is just a destructive interference, which results in the revival's damping.
\subsection{\label{subsec:dec3} Three-body losses}
The main source of three-body losses is the recombination of condensed atoms, discussed in details in both theoretical \cite{moerdijk1996, fedichev1996} and experimental \cite{burt1991, soding1999} works.
Our study of three-body losses is just a simple extension of the previous subsection. We propose, also in phenomenological manner, the following master equation
\begin{eqnarray}
\nonumber\frac{d\hat{\rho}}{dt}&=&-\frac{i}{\hbar }\left[\hat{H}_A, \hat{\rho }\right]+\kappa_3 \sum_{k=1}^2  \left[ \left(\hat{a}_k\right)^3, \hat{\rho } \left(\hat{a}_ {k}^{\dagger }\right)^3\right]\\
& &+\left[ \left(\hat{a}_k\right)^3 \hat{\rho}, \left(\hat{a}_ {k}^{\dagger }\right)^3\right] ,
\label{eqn:master3}
\end{eqnarray}
where $\kappa_3$ is in the relation to the recombination event rate constant $K_3$ (see for instance \cite{sinatra1998, jack2002, gerton1999})
\begin{displaymath}
\kappa_3 = \frac{K_3}{6} \int \left|\psi (\bm{r})\right| ^6 \mbox{d}^3r.
\end{displaymath}
Like in the previous case -- the equation above was derived in the limit of suppressed tunneling between wells. The results of the master equation in our case, namely $J > 0$, we treat only as a qualitative estimation.

The non-hermitian evolution operator and quantum jumps operator have forms
\begin{eqnarray}
\nonumber\tilde{H}&=& \hat{H}-\dot{\imath} \hbar \kappa_3 \left(\left(\hat{a}_1^{\dagger}\right)^3\hat{a}_1^3+\left(\hat{a}_2^{\dagger}\right)^3\hat{a}_2^3\right)\\
\nonumber\hat{C}_1&=& \sqrt{2\kappa_3} \hat{a}_1\\
\hat{C}_2&=& \sqrt{2\kappa_3} \hat{a}_2.
\label{eqn:jumps3}
\end{eqnarray}

\begin{figure}
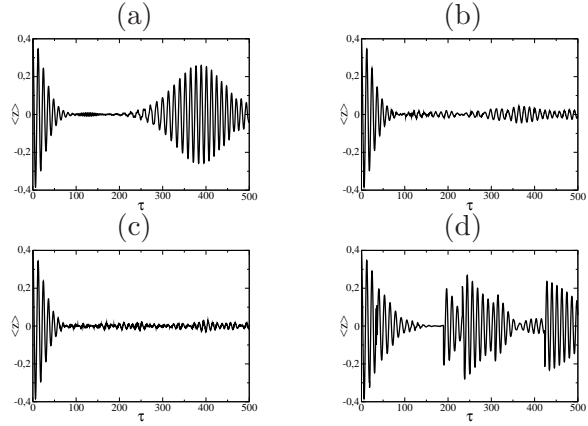

\begin{minipage}[t]{0.20\textwidth}
\centering
(a)

\includegraphics[width= \textwidth, clip]{exact3}
\end{minipage}
\hspace{0.05\textwidth}
\begin{minipage}[t]{0.20\textwidth}
\centering
(b)

\includegraphics[width= \textwidth, clip]{z3}
\end{minipage}

\begin{minipage}[b]{0.20\textwidth}
\centering
(c)

\includegraphics[width= \textwidth, clip]{z6}
\end{minipage}
\hspace{0.05\textwidth}
\begin{minipage}[b]{0.20\textwidth}
\centering
(d)

\includegraphics[width= \textwidth, clip]{z3_3body_single}
\end{minipage}
\caption{Evolution of $\langle z \rangle (\tau)$ (a) without losses, (b) with in average one quantum jump before revival, (c) with in average four quantum jumps before revival. (d) Single realization of stochastic wave approximation. In the all figures $z_0=0.4$, $\phi_0=0$,  the initial total number of atoms $N=1000$ and $\gamma=1$.}
\label{fig:3body}
\end{figure}

We present results of numerical simulations in Fig. \ref{fig:3body}. Figures (a), (b) and (c) display the evolution up to the first revival without losses, with on average one quantum jump and with on average four quantum jumps, respectively. Like in the previous subsection the revivals are strongly suppressed even with relatively small losses and a single realization (Fig. \ref{fig:3body}(d) of stochastic process differ from the average evolution.

Assuming that to observe the phenomenon, losses should be smaller than one atom until the revival, we get the qualitative criterion for $\kappa_3$
\begin{displaymath}
\kappa_3 N^3 t_{rev}<1.
\end{displaymath}
For example for $N=1000$ and $t_{rev}=50 \mu s$ one get $\kappa_3 < 0.5\times10^{-10}$/s. From the rough estimation of the possible experimental setups ($K_3<10^{-28}$cm$^3$/s (\cite{gerton1999, esry1999}) and the size of each BEC in the wells about few micrometers) we compute the parameter $\kappa_3$ of the same order of magnitude. 
\section{\label{sec:concl}Conclusions}
In this paper we investigated  attractive BEC in a double-well potential. We first recognized three essential time scales characteristic for the evolution of the relative number of atoms' imbalance between the two potential wells. The first timescale, the Rabi oscillation, has a purely classical origin. It can be understand within the frame of the mean field approximation. The second timescale is the time of collapse. We have explained it within the semiclassical picture, using differential equation from the mean field approximation but for non-deterministic initial values. Eventually we explain the phenomenon of revivals -- this turns out to be exactly the same as for revival in the Jaynes-Cummings model. However we also show that the phenomenon does not require the quantum model. It may be also understood using theoretical tools of classical physics. This means that these revivals are not the evidence of quantum correlation in the system.

We investigated also sensitivity of revivals for one- and three-body losses. We showed that even if the fraction of the lost number is small (i. e. $0.1$\%), the revival is strongly suppressed. On the other hand the rate of one-body losses in the recent experiments is so small, that this kind of losses should be negligible. On the contrary, the three-body losses might be an important limitation. We do not exclude the revivals from being measured but our results indicate that experiments should be prepared for relatively low densities.
\begin{acknowledgments}
\label{sec:acknowl}
The work was supported by Polish Government research funds for 2009-2011 (K. R. and K. P.) and 2007-2010 (M.T. and P. Z.)
\end{acknowledgments}

\end{document}